# Three-Tone Intermodulation Distortion Generated by Superconducting Bandpass Filters

Stephen K. Remillard, *Member, IEEE*, H.R. Yi *Senior Member, IEEE*, and Amr Abdelmonem, *Senior Member, IEEE*

***Abstract*— Microwave bandpass filters constructed from materials exhibiting some nonlinearity, such as superconductors, will generate intermodulation distortion (IMD) when subjected to signals at more than one frequency. In commercial applications of superconductive receive filters, it is possible for IMD to be generated when a weak receive signal mixes with very strong out-of-band signals, such as those coming from the transmitter. A measurement procedure was developed and data were taken on several different types of superconducting bandpass filters, all developed for commercial application. It was found that in certain interference situations, the 3-tone mixing can produce a spur that is noticeable by the receiver, but that there are simple preventative design solutions.**

**Index Terms— High Temperature Superconductors (HTSs), Filters, Intermodulation Distortion (IMD).**

## I. INTRODUCTION

By now, the application of HTS filters to wireless telecommunications has been well documented [1]. A wireless receiver benefits from inserting an HTS filter between the antenna and the receiver in two key ways. Being of high order, the HTS filter offers sharp rejection of out-of-band interferers. Being of high Q and low temperature, the HTS filter does not desensitize the receiver with added noise figure nearly as much as a conventional filter does. However, the HTS material is not perfectly linear, and it is possible in some situations for the HTS filter to actually degrade the front-end over the performance with a conventional, non-HTS, filter by introducing intermodulation distortion (IMD). This paper identifies the critical IMD measurement, presents results on commercial HTS filters, uses RF system engineering to determine an acceptable upper limit, and discusses filter design solutions to meet that system requirement.

S.K. Remillard was with ISCO International, Mount Prospect, IL 60056 USA, and is now with Agile Devices, Inc., Evanston, IL 60201 USA (phone 847-570-4932; e-mail: Remillard@ieee.org).

H.R. Yi was with ISCO International and is now with Netcom, Inc., Wheeling, IL 60090 USA.

A. Abdelmonem is with ISCO International.

## II. Overview of the Filters Used in this Study

HTS bandpass filters are being routinely made using two significantly different technologies. In one case, filters are fabricated using, for example, thin film $YBa_2Cu_3O_{7-\delta}$ (YBCO) on MgO substrates, or using thin film $Tl_2Ba_2CaCu_2O_8$ on $LaAl_2O_3$ substrates. The compact resonator design is described elsewhere [2]. The nonlinearity of a 1.5 MHz wide 8-pole filter at 845 MHz made with YBCO on MgO will be reported below. This filter is targeted for the A-prime cellular band. Also, the nonlinearity of a 7.5 MHz 8-pole filter at 900 MHz made from folded hairpin resonators using TBCCO will be included [3]. This filter is targeted for GSM-900 base stations.

In the other case, three-dimensional filters are made by applying thick film coatings of $YBa_2Cu_3O_{7-\delta}$ to yttria stabilized zirconia (YSZ) substrates, or to silver plated stainless steel substrates. A 20 MHz wide 16-pole filter at 1950 MHz used compact closed-slot spiral resonators made from YBCO on YSZ substrates [4]. This filter is targeted at the 3rd generation wireless band now being used in Japan. A 1.5 MHz wide 8-pole filter at 845 MHz was made from YBCO applied to a split toroidal resonator [5]. Like the 8-pole thin film filter above, this filter is also targeted to the A-prime cellular band, and the two will be compared. A 9.5 MHz wide 14-pole filter at 895 MHz was made from YBCO applied to metallic quarter wave resonators in a comb-line configuration. Like the 7.5 MHz wide 8-pole thin film filter above, this filter is targeted to the GSM-900 band, and the two will be compared.

## III. Three-Tone IMD Measurement

### *A. The origin of 3-Tone IMD*

Intermodulation occurs when signals at more than one frequency are incident on a nonlinear device as discussed in [6]. A special case is often made of IMD occurring from the mixing of two signals. This 2-tone IMD is especially convenient for semiconductor devices because it provides a standard test condition for device characterization. It is perhaps more simplistic than the real device application, since more than two signals, or tones, are usually incident. Nevertheless, the 2-tone IMD conveniently indicates the extent of a device's nonlinearity.

Like semiconductor devices, an HTS filter can also be characterized by measuring the IMD generated by two signals passing through it, but that does not indicate the impact of the IMD on system performance. Filters are frequency selective, and the nonlinear performance is not the same for in-band and for out-of-band signals. Some authors (including the present [7]) continue to rely on 2-tone IMD from in-band carriers, as a convenient way to characterize and to compare HTS filters. But, within the context of an application, the in-band signals are small enough (typically <-50 dBm) that the HTS filter does not generate IMD. However, out-of-band signals are usually much stronger (as strong as +15 dBm), creating a real nonlinear output from the filter. Two strong out-of-band signals by themselves usually do not pose a threat because the order of the IMD they can generate in-band is high. However, a real threat to the radio receiver occurs when two closely spaced out-of-band signals mix with one in-band signal. Then it does not matter how far out-of-band the strong carriers are, they will provoke a nonlinear output within the filter passband.

When three tones are incident on a nonlinear device, the mixing product of greatest concern is that which occurs nearest to the desired signal. This mixing term is predicted in the same way as the mixing terms from 2-tone intermodulation, as described in [6] and [8]. Harmonic terms are found from a Taylor series expansion of the electric field

$$E = E(0) + \left.\frac{dE}{dH}\right|_{H=0} H + \left.\frac{d^2E}{dH^2}\right|_{H=0} \frac{H^2}{2} + \left.\frac{d^3E}{dH^3}\right|_{H=0} \frac{H^3}{6} + \cdots \quad (1)$$

where the magnetic field, H, contains three input frequencies, $H_{in}=H_1\sin(\omega_1 t)+H_2\sin(\omega_2 t)+H_3\sin(\omega_3 t)$, where $\omega=2\pi f$. When the magnetic field is inserted into Equation (1) there are four terms, all from the third order expansion term, which include all three input frequencies. IMD products are expected at any positive values of $f_1-f_2+f_3$, $f_1-f_2-f_3$, $f_1+f_2+f_3$, and $f_1+f_2-f_3$. If $f_1$ and $f_2$ are the frequencies of two out-of-band signals, and $f_3$ is the frequency of an in-band signal, then a third order IMD product will occur at $f_{IMD}=f_3\pm(f_2-f_1)$ and is described by

$$E_{IMD} \propto E_1 E_2 E_3 \sin[(\omega_3 \pm (\omega_1 - \omega_2))t] \quad . \quad (2)$$

$E_{IMD}$, which can be regarded as electric field or voltage, depends linearly on the amplitude of each carrier signal. If one signal changes by a factor of 10, then $E_{IMD}$ should change a factor of 10. If all three change by a factor of 10, then $E_{IMD}$ should change a factor of 1,000. Measurements performed on a GaAs low noise amplifier using the circuit described below yielded exactly that result. When the IMD was plotted on a logarithmic scale (in dB) versus the input power in dB, the slope was 1:1 when only one signal varied, 2:1 when two signals varied identically, and 3:1 when three signals varied identically.

*B. Measurement Circuit*

In order to measure 3-tone IMD, a circuit must be constructed to isolate and combine the three signals of interest. The circuit used by the authors is shown in Figure 1. Three high dynamic range signal generators produce the test signals. The out-of-band signals are first combined together by a 3 dB hybrid coupler. A 2:1 power combiner can be used in place of the splitter. An optional step attenuator can be used to adjust the signal level. The power control on the signal generators can also be used instead of the step attenuator. A power amplifier (PA) then boosts the two signals. The PA will generate IMD, but as long as the two signals are far enough out of band that the PA's IMD is of very high order, then it will not affect the measurement. In cases where the out-of-band signals are close to band, such that the PA's IMD at $f_{IMD}$ is less than approximately 10$^{th}$ order, then two separate PAs should be used before the hybrid.

The isolator catches the reflected carriers $f_1$ and $f_2$, which dissipate in the load on its 3$^{rd}$ port. The isolator, combined with the notch filter and the optional 3 dB attenuator, also isolates the PA output stage from the third carrier, $f_3$. This isolation is necessary since the hybrid (of particular concern, the one on the right) only provides about 20 dB of isolation between the input ports. The notch filter is tuned to $f_3$ and is usually about 20 dB deep. The bandpass filter on the path that generates $f_3$ isolates the signal generator from $f_1$ and $f_2$, which can enter this path either through the imperfect isolation of the hybrid, or as reflections from the HTS filter. The bandpass filer must provide at least 30 dB of isolation at $f_1$ and $f_2$. The hybrid coupler on the right combines all three carriers, which then impinge upon the HTS filter. An optional bandpass filter after the HTS filter provides further isolation of the spectrum analyzer from bleed through of carriers $f_1$ and $f_2$.

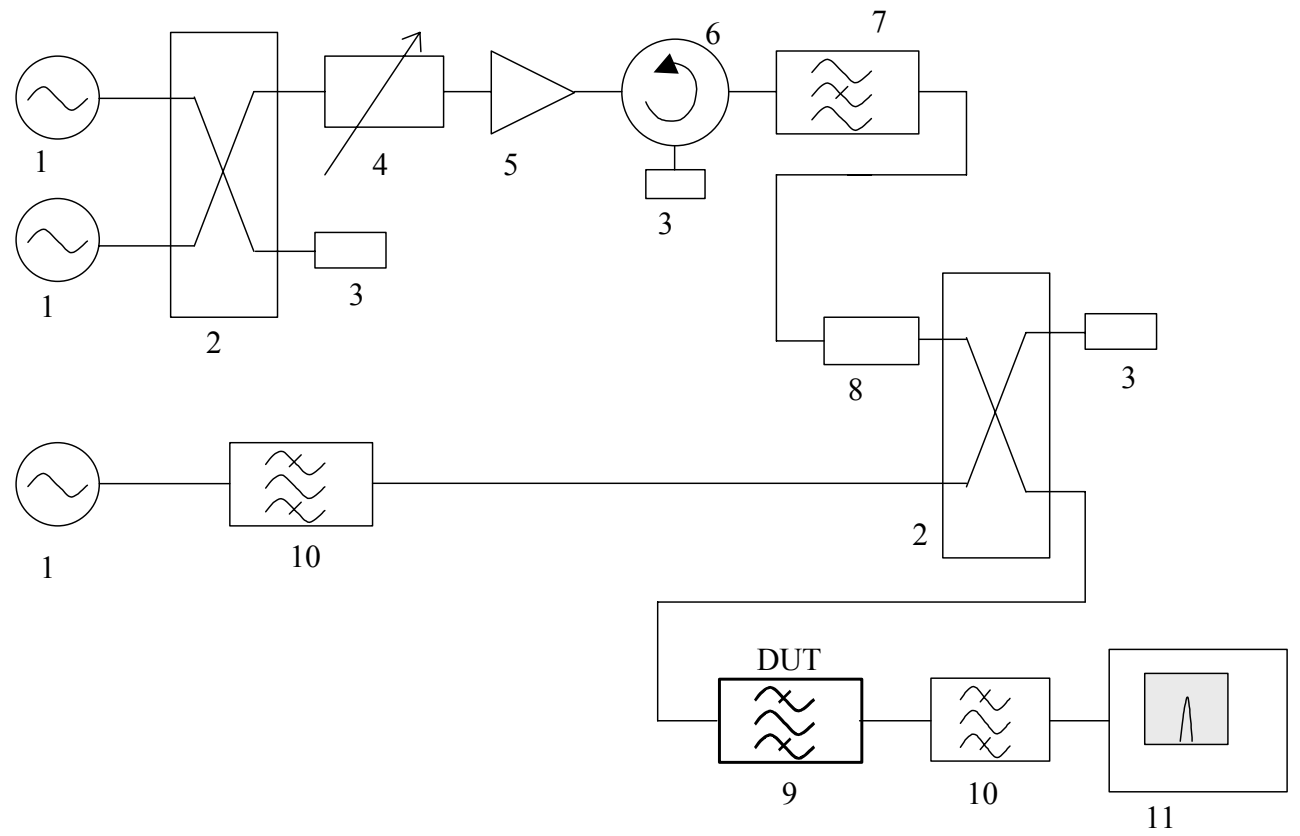

Fig. 1. The circuit used to measure 3-tone IMD. 1. Signal generator; 2. Hybrid coupler; 3. 50 Ω load; 4. Step attenuator; 5. Power amplifier; 6. Circulator; 7. Notch filter; 8. 3 dB attenuator; 9. HTS filter; 10. Added protection bandpass filter; 11. Spectrum analyzer.

Before measuring the 3-tone IMD with the HTS filter, a conventional filter that passes $f_3$ and rejects $f_1$ and $f_2$ should be inserted in place of the DUT. Steps sometimes need to be taken to reduce any "set-up IMD" found this way to a level at least 20 dB smaller than the IMD measured with the HTS filter. In all measurements in this work, the set-up IMD was at least 15 dB lower than the DUT, and usually undetectable. For example, when $f_1$ and $f_2$ were +20 dBm and $f_3$ was -10 dBm, the set-up IMD was -118 dBm. As a worst case for set-up IMD interfering with measured IMD, the A-prime band thick film HTS filter, which had lower IMD than the other filters, had about -97 dBm at 77K in this case. When $f_3$ was lowered to -20 dBm, the set-up IMD could not be measured with the sensitivity of the equipment (about -130 dBm). The A-prime band thick film filter measured –108 dBm in this case. These set-up IMD levels give an indication of the sensitivity of this measurement. For very low level carrier signals, the sensitivity of the apparatus can be improved from about –130 dBm to about -150 dBm by attaching a low noise amplifier to the input port of the spectrum analyzer.

## IV. Measurement Results

### *A. Two-Tone Results*

Two-tone IMD measurements are well documented [7]. In a two-tone measurement, two carriers, $f_1$ and $f_2$, either in-band or out-of-band of the filter, are incident on the filter. The third order IMD spurs are measured at $2f_1$-$f_2$ and at $2f_2$-$f_1$. In the case of studies of the 2-tone IMD from HTS filters, certain observations have been made. (1) The IMD improves as temperature is reduced. Below $T_C/2$, actually below about 60 Kelvin for YBCO, there is very little temperature dependence, as has been seen in our own data and is evident in published data as well [9]. (2) For carriers near the band-edge, especially carriers in the filter skirts, the higher group delay in that frequency region results in higher currents relative to the middle of the band [10]. This also results in higher IMD for carriers in this region.

When 3rd order 2-tone IMD is plotted logarithmically, in dBm versus input power in dBm, for a semiconductor device, the slope is usually 3:1. For HTS filters, the slope is determined by the extent to which the surface impedance, $Z_S(H_{RF})$, depends on surface RF magnetic field, $H_{RF}$. The IMD (and hence its slope) of thin film microstrip filters is generated by currents contained within one penetration depth of the film edge [11]. Three dimensional thick film HTS coated structures usually have no current-carrying edges, which significantly reduces the kinetic inductance contribution. Their surfaces are comprised of superconductive grains, which have weak electrodynamic coupling. The IMD, and especially its slope, depends on the RF current dependence of the surface resistance of the HTS thick film [8], which in-turn depends on inter-granular coupling. The power law field dependence of $Z_S$ on $H_{RF}$ varies across different magnitudes of field, leading to changes in the slope of the IMD tone at different carrier power levels. For several reasons treated in Chapter 3 of [10], the slope is found to vary from 1:1 for very granular HTS material to 2.5:1 (or even nearly 3:1) for high quality epitaxial material.

### *B. Three-Tone IMD Studies Made with Different Filters*

#### *1) GSM Band Filters*

Three-tone IMD measurements were made on the two 900 MHz filters. The measured IMD for the 7.5 MHz wide filter made from TBCCO thin film on LAO is shown in Figure 2. The major portion of each curve has a slope of almost exactly 1:1 over the power range of the measurement, even though two signals are being varied, leading one to expect a slope of 2:1 per (2). The slope of the IMD when only the in-band signal level is varied was also nearly 1:1, as expected from (2).

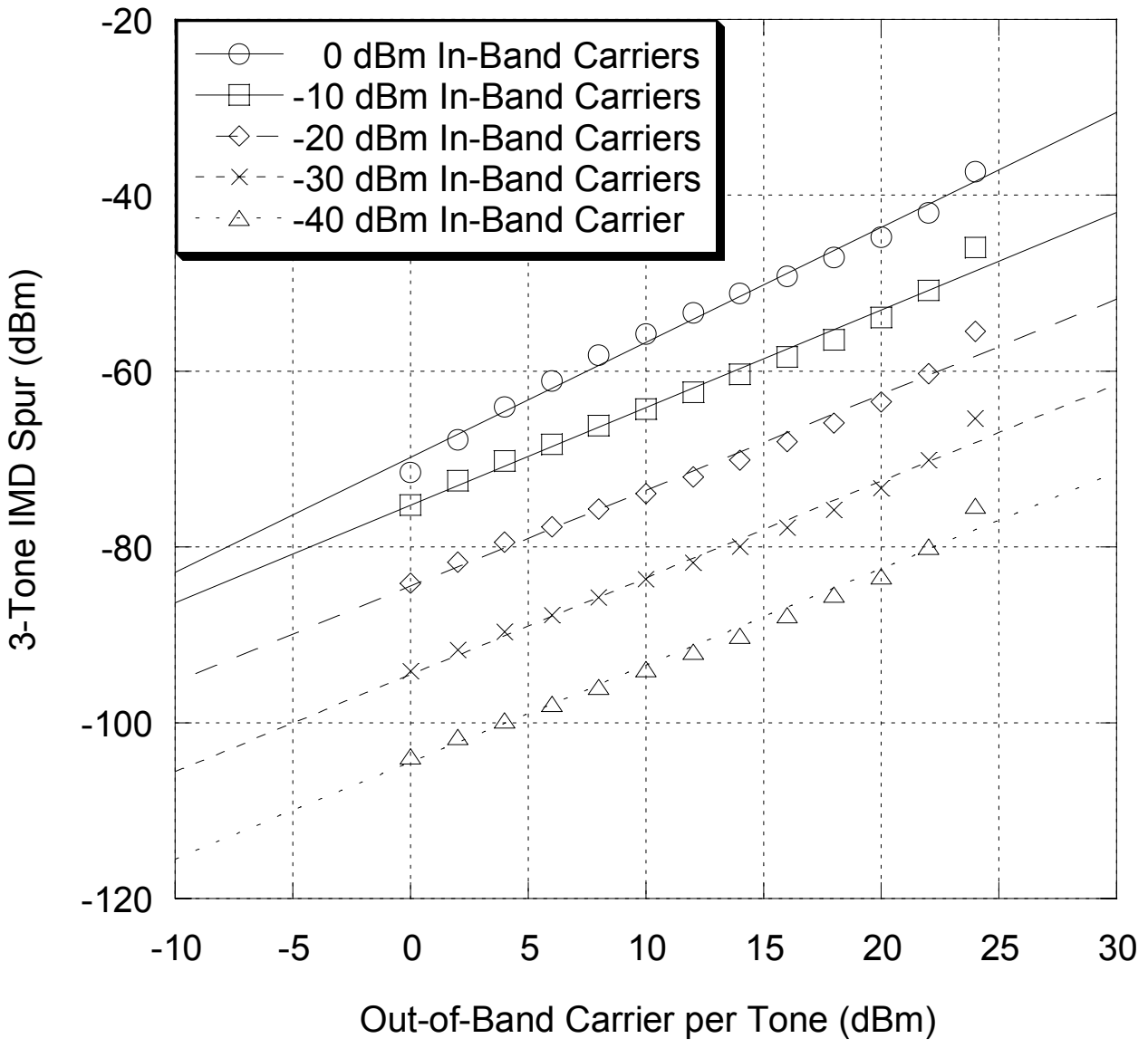

Fig. 2. The 3-tone IMD measured for an 8-pole, 7.5 MHz wide filter centered at 903.75 MHz fabricated from $Tl_2BaCa_2Cu_2O_8$ thin film on a Lanthanum Aluminate (LAO) substrate was measured at 75 Kelvin. The two out-of-band carriers were spaced 0.5 MHz apart.

The 9.5 MHz wide filter was made from YBCO thick film on Ag-plated, stainless steel three-dimensional substrates and any 3-tone IMD was below the sensitivity of measurement. The input and the output resonators of this 14-pole filter were not superconducting. Because the currents from out-of-band carriers reside predominantly in the input resonator [12], it is primarily in the input resonator that the IMD from three-tone mixing is generated. The practice of using the bare metallic substrates for the I/O resonators does not significantly degrade the filter performance [13], except for very narrow filters.

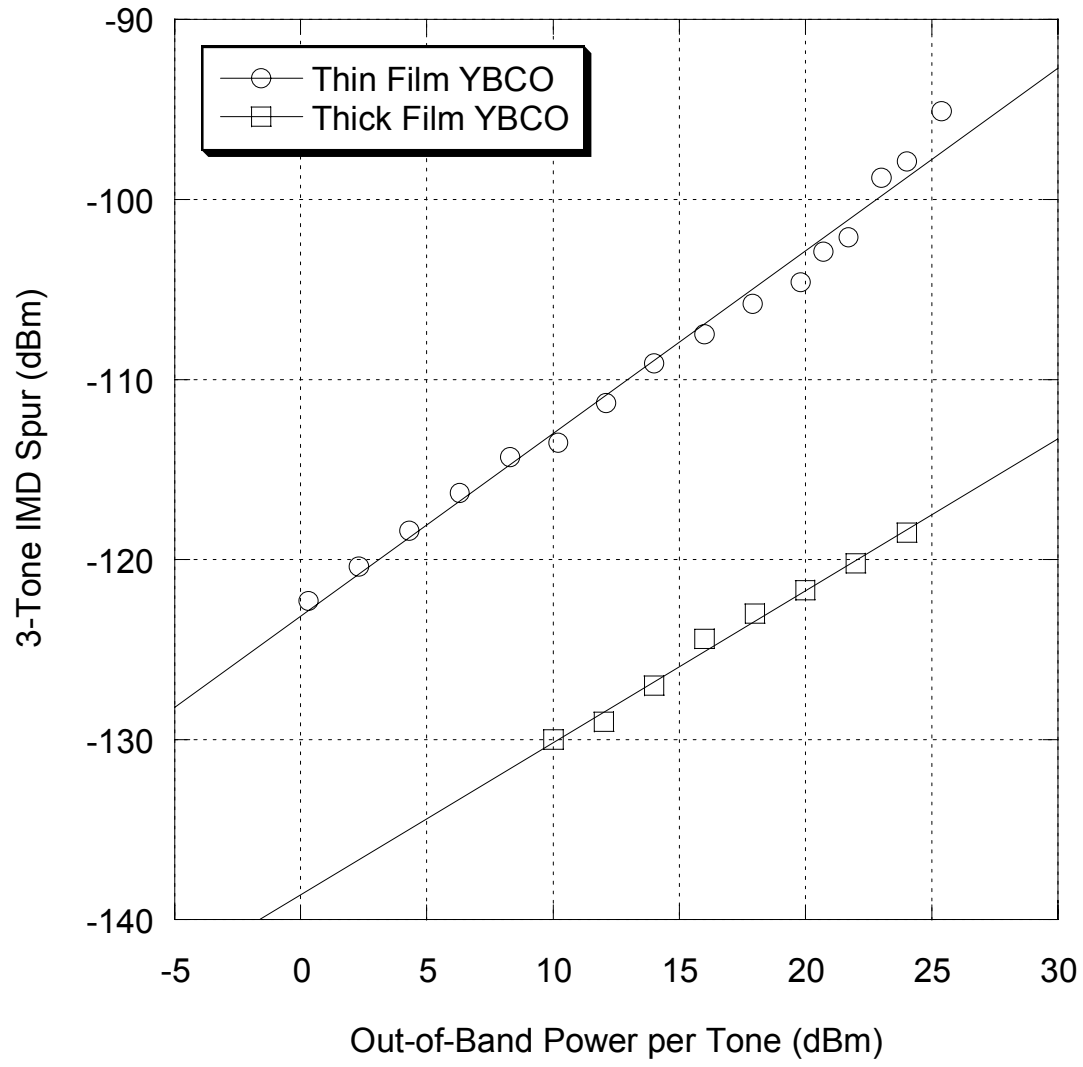

Fig. 3. Two 8-pole, 1.5 MHz wide filters centered nominally at 845 MHz were made, one from thick film YBCO and one from thin film YBCO on an MgO substrate. The IM of the two at 72.5 Kelvin is compared here. The out-of-band carriers were at 870 MHz with 1 MHz separation. The in-band carrier was at -30 dBm

*2) A-Prime Band Filters*

Three-tone IMD measurements were made with the two YBCO 1.5 MHz wide 8-pole filters nominally tuned to 845 MHz. These are shown for a temperature of 72.5 Kelvin and a –30 dBm in-band carrier in Figure 3. The IMD is higher for the thin film filter. However, because the I/O resonators in the thick film filter were superconductive, the three-tone IMD was large enough to measure. In the case of three-tone IMD, the slopes of the IMD spur either versus out-of-band power or versus in-band power are nearly 1:1.

*3) 3G/UMTS Band Filter*

Third generation Universal Mobile Telecommunications System wireless was launched in Japan in 2001, and is now being launched in parts of Europe. A highly compacted three-dimensional thick film YBCO coated resonator was developed for making filters for this service at 1950 MHz that are a factor of 6 smaller than the previous generation thick film YBCO filter. Three-tone IMD was measured under two different conditions with this 16-pole, closed slot spiral resonator filter. The first condition was with the out-of-band carriers positioned in the assigned transmit band, 180 MHz away. The second condition was made with the out-of-band carriers very near to the band-edge, just 2.5 MHz away. It can be seen that the IMD is much higher when the carriers are closer to the passband.

A more complicated circuit is needed to perform the second measurement. In this case, separate power amplifiers are used for each signal path since passing both out-of-band tones through one PA will produce significant IMD in the filter passband. Figure 4 shows one curve for each case. The IMD is 10 to 15 dB higher for the close-in carriers, despite the more than 80 dB of rejection offered by the HTS filter 2.5 MHz out of band.

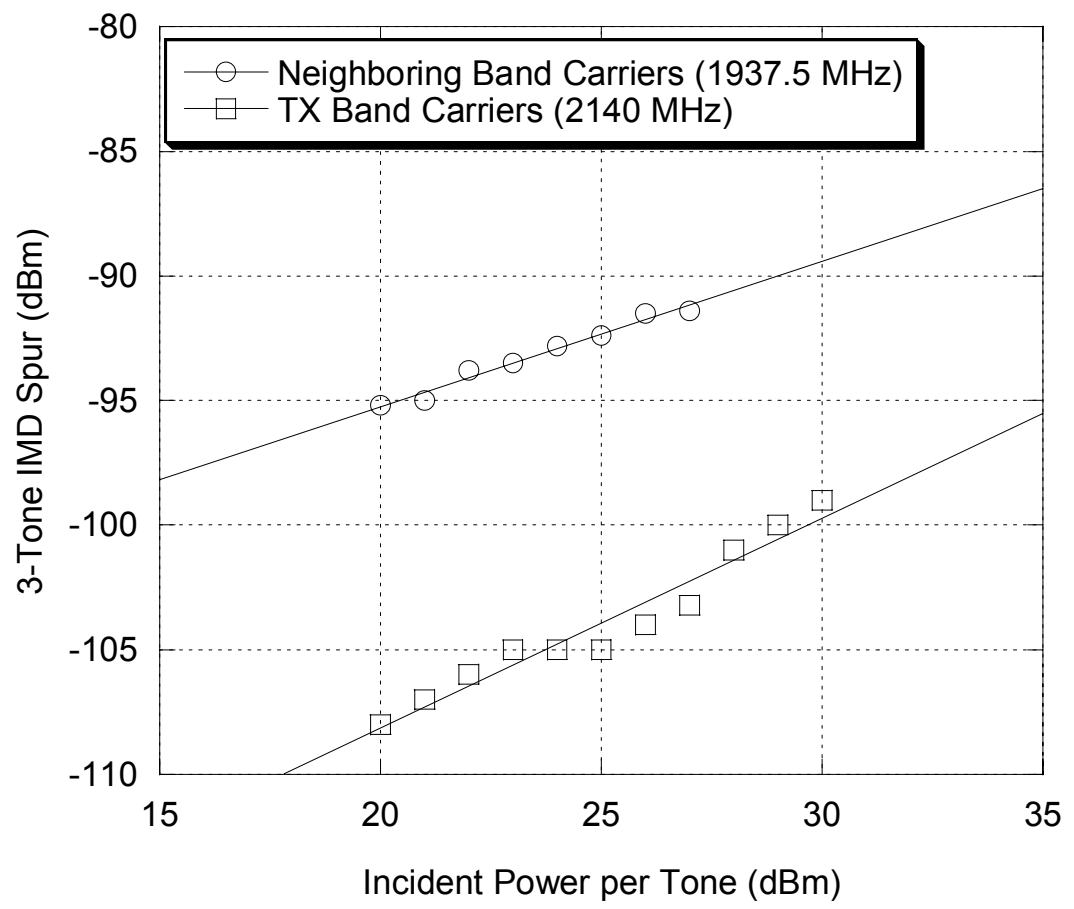

Fig. 4. 3-tone IMD was measured on a 20 MHz wide 16-pole filter centered at 1950 MHz. In one measurement, the out-of-band carriers were 180 MHz away from the passband at 2139.5 and 2140.5 MHz, where filter rejection was >130 dB. In the other they were 2.05 MHz and 2.65 MHz out of band, where filter rejection was about 85 dB. The in-band carrier was -30 dBm at 1950 MHz.

## C. Behavior of 3-Tone IMD

*1) Dependence of IMD on the Frequency of the Out-of-Band Carriers*

The result in IV.B.3 showing the elevated IMD level that occurs when the carriers are in the neighboring band, instead of in the more distant TX band, motivated further investigation of the frequency dependence. From the two-tone result that IMD is higher when the two in-band carriers are close to the band edge, it is expected that higher three-tone IMD occurs when the out-of-band carriers are located in the filter skirt, or near a transmission zero. This was indeed the observation in IV.B.3.

The IMD versus the frequency of the out-of-band carriers was measured for the 1.5 MHz wide thick film HTS filter. The out-of-band carrier frequencies were varied from 857 MHz to 905 MHz. The measured IMD at each frequency versus the rejection of the filter at each frequency is plotted in Figure 5. Although, no physical connection between IMD and rejection is postulated here, the correlation between the IMD and the rejection of this filter is about 1 dB of added IMD for every 4 dB of less rejection.

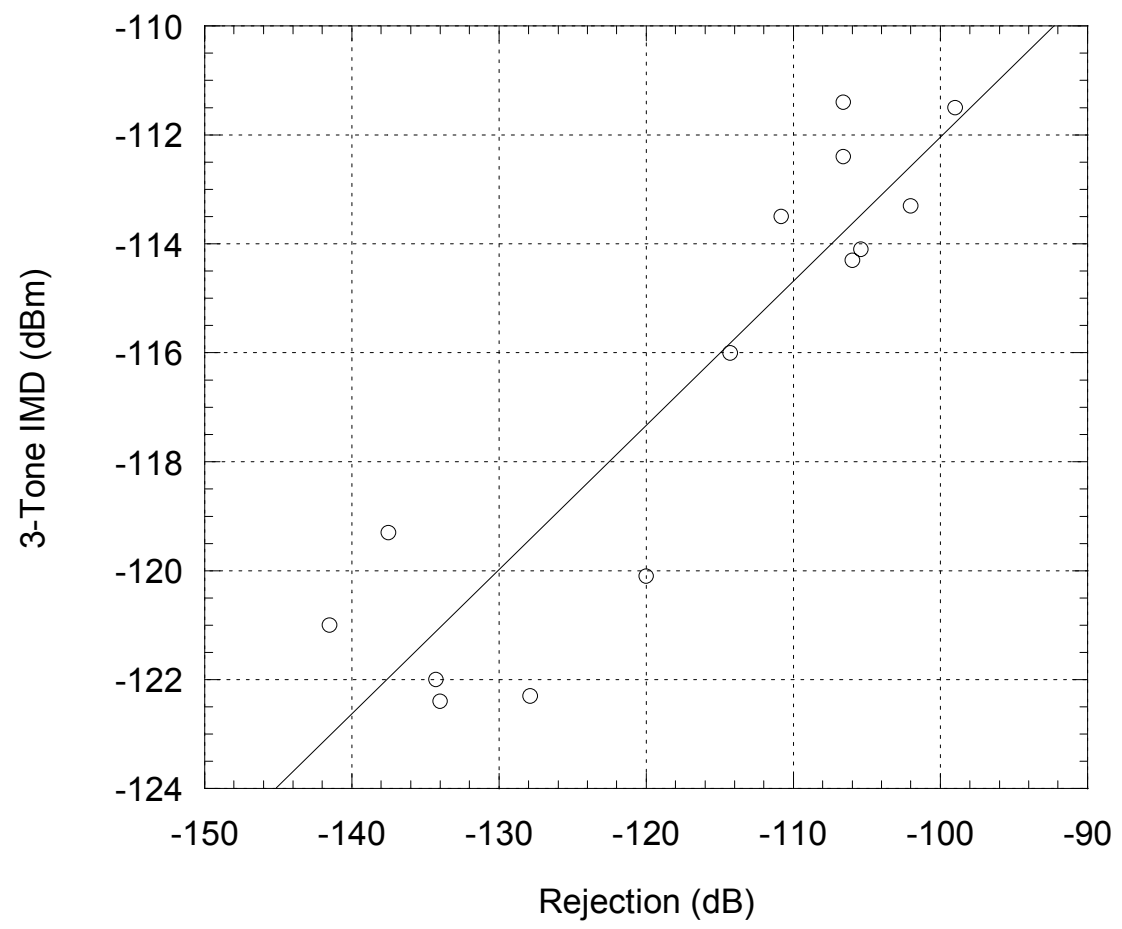

Fig. 5. The variation of 3-tone IMD with filter rejection was measured using the thick film A-prime band filter. Varied rejection was achieved by performing the measurement with the 1 MHz spaced out-of-band carriers at different frequencies.

*2) Temperature Dependence of the Three-Tone IMD*

The nonlinear effects of superconductors peak just below $T_C$. This has been shown by several groups to be the case in passive HTS microwave devices [14] and [15]. The temperature dependence of the 3-tone IMD of the thick film A-prime band filter is shown in Figure 6. Thick film superconductors are very granular, and as the temperature reaches $T_C$ (92 Kelvin) the current needed to break down the inter-granular Josephson coupling between grains (usually referred to as $I_{CJ}$) becomes smaller, resulting in higher nonlinear distortion.

## D. Measurement Reproducibility

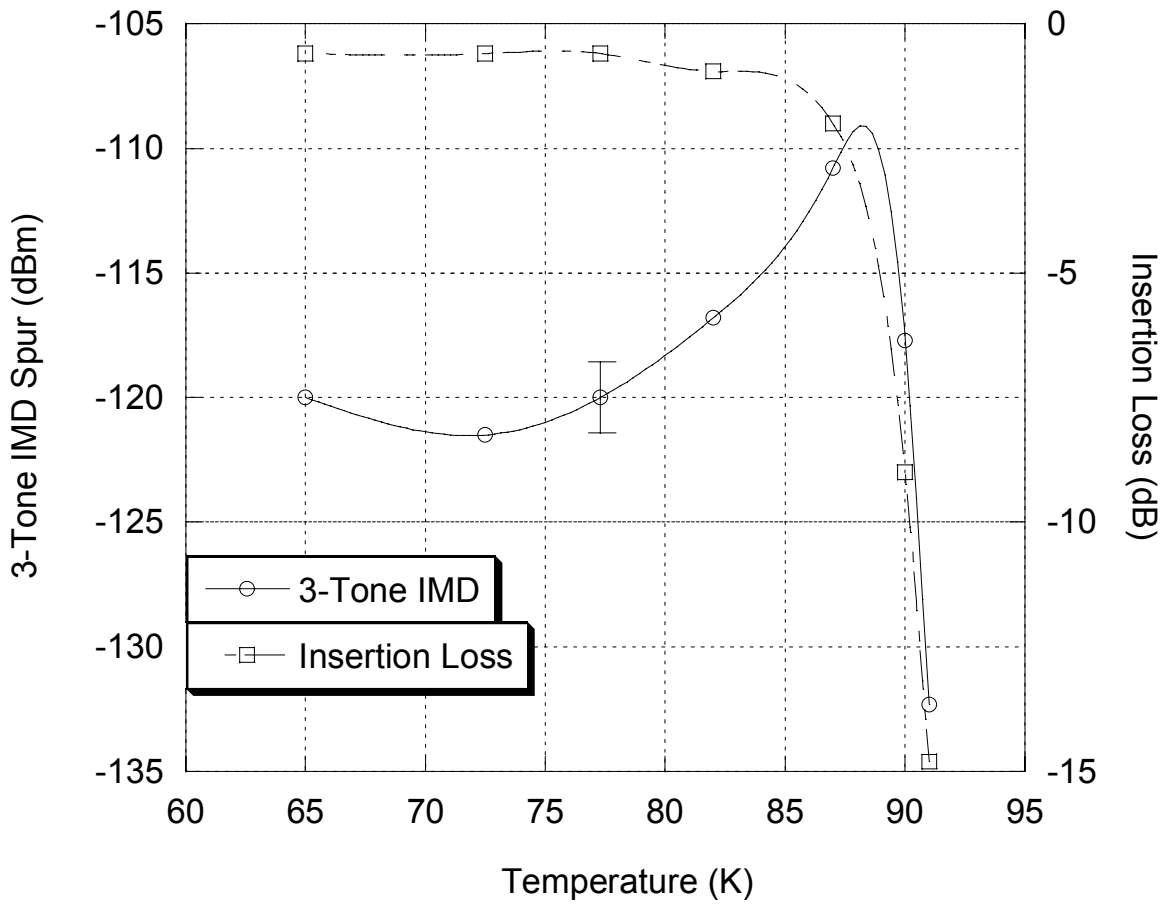

Fig. 6. Cubic spline fit to the 3-tone IMD at various temperatures of a bandpass filter showing the characteristic peak in nonlinearity close to $T_c$, with a sudden drop in nonlinearity during the transition.

In the case of these measurements the Hewlett-Packard signal generators and spectrum analyzer accuracies were validated through recent factory certified calibrations, so the uncertainty lay in the set-up. The insertion loss or gain of each component was measured using a network analyzer, also with a recent factory certified calibration. The accuracy then with which signal levels are generated and detected is well within ±0.5 dB. So, the best assessment of the measurement is through its reproducibility. Figure 5 gave an indication of reproducibility through the scatter of the data about the straight line. Each data point represents data from a separate set of IMD measurements, occasionally interrupted by re-tuning the notch filter in the circuit of Figure 1. The average deviation of data about the line is ±1.44 dB. As a separate consideration, the 3-tone IMD of the thin film A-prime band filter was measured at 75 Kelvin with +14 dBm out-of-band carriers and –10 dBm in-band carrier on three different occasions, with the test circuit having been re-constructed each time. The three values were -89.9 dBm, –88.4 dBm and –91.6 dBm, or –90.0±1.6 dBm where the uncertainty of 1.6 dB is the standard deviation. We will consider ±1.6 dB to be the reproducibility of these 3-tone IMD measurements.

## V. Application to Wireless Communication

HTS filters are used in the front-end of radio receivers immediately after the antenna. They are used to reduce interference and, because of the low temperature of operation, to improve sensitivity. The data in the previous section demonstrate that the filters can inject interfering signals in the form of IMD. This section will answer the question of what level of IMD is unacceptable, and describe solutions to ensure the IMD is at an acceptable level.

The data in IV show that strong out-of-band carriers impinging on HTS filters cause unwanted in-band interference when mixing with an in-band carrier. The data indicate an operational limit for the filter in question. However, the operational requirements of HTS filters need to be defined. HTS filters are used in cellular base stations that operate within the 800 MHz cellular (824-849 MHz), GSM 800 (890-915 MHz), 3G (1920-1980 MHz) and PCS (1850-1910 MHz) receive bands. Each operator uses a sub-band within these bands, and the receiver is best served by HTS when the HTS filter response conforms to the sub-band.

For each base station receive band there is a corresponding base station transmit band. The transmit band for 800 MHz cellular is 869-894 MHz. In base station applications, high energy carriers from the transmitter usually impose the strongest out-of-band signals on a receive filter. Figure 7 shows a spectrum measurement of the energy received by the antenna of an 800 MHz cellular base station. The measurement was taken at the bottom of the antenna feed cable, where it goes next to the HTS filter. Figure 7a shows the spectrum between 860 MHz and 900 MHz. This includes the base station's transmitter at 880-894 MHz. The out-of-band interferers are as large as +15 dBm, and are as close together as 1.3 MHz. To be fair, this is one of the worst examples of transmit band interference, which is usually closer to 0 dBm. Figure 7b shows the in-band energy for this busy Time Domain Multiple Access base station. The largest in-band peak is about –65 dBm. If the base station filter has the IMD performance of the filter from Figure 2, then an extrapolation would indicate that the IMD level will be about –112 dBm.

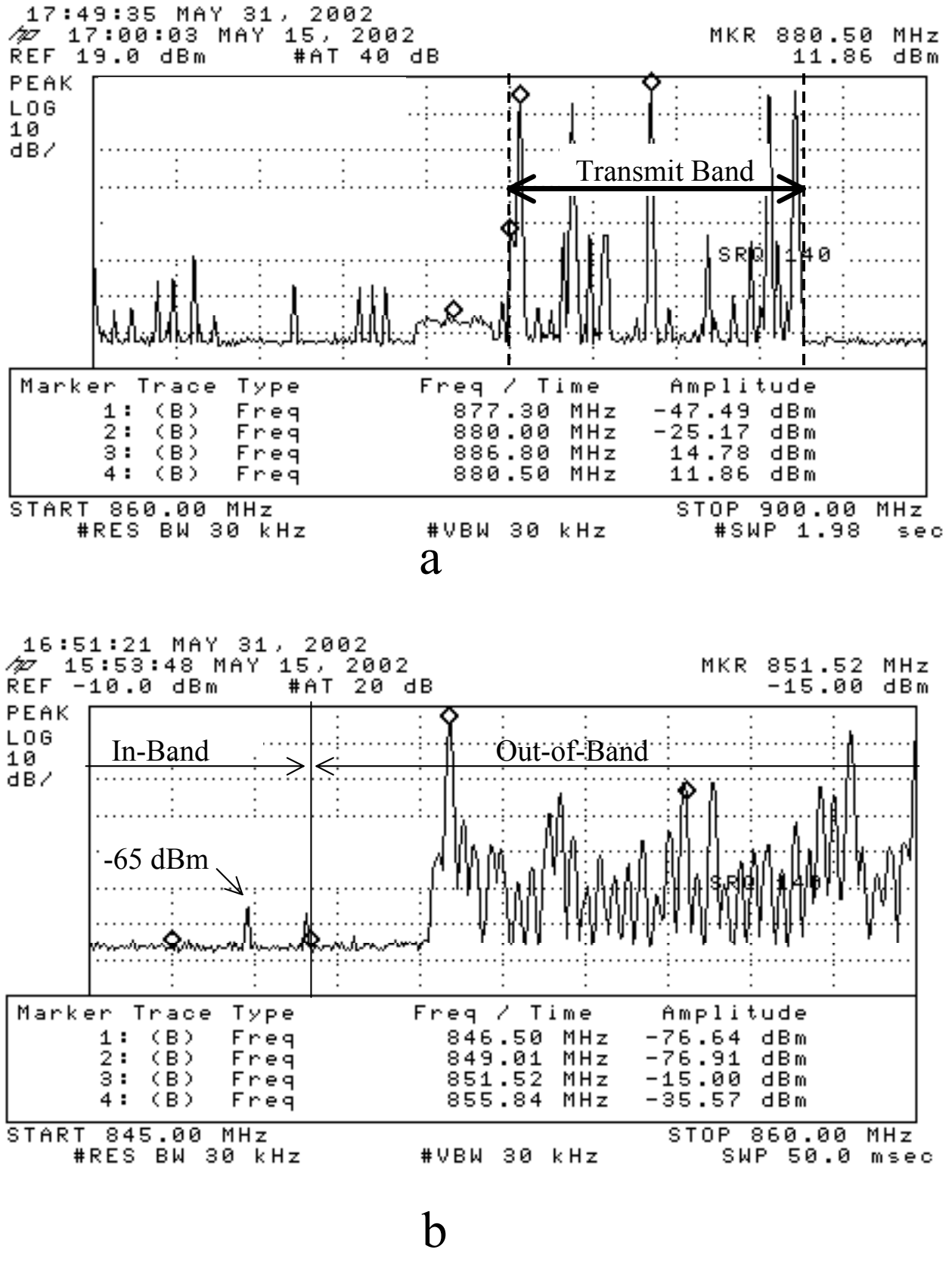

Fig. 7. Spectrum analyzer sweeps taken from the receive antenna at an 800 MHz cellular B-Band base station showing (a) the spectrum containing the transmit band, and (b) the spectrum containing the desired in-band energy.

The maximum acceptable level of in-band interference is a complicated function of the receiver characteristics. However, if purveyors of HTS filters do not wish IMD to be an issue with the end-user, then the IMD of the HTS filter should be at least 10 dB below the receiver sensitivity, which is the minimum signal level at the receiver input needed to maintain a required receiver performance metric. In the case of IS-95 Code Division Multiple Access, the receiver metric is a frame error rate of 1%, and a reasonable receiver sensitivity level is –117 dBm [16]. The data in this paper demonstrate that some HTS filters, at least the filter in Figure 2, are likely to introduce visible distortion to the radio receiver.

There are practical solutions to the problem of 3-tone IMD. The reader is now reminded of the solution used in the YBCO thick film GSM band filter in IV.B.1 to realize no 3-tone spurs, where the input resonator was not superconducting. High input resonator Q is unnecessary in all but the most extremely narrow band filters. Another means of stopping 3-tone IMD is to place a low order (perhaps 3-pole) filter in front of the HTS filter, at the expense of added insertion loss. The filter designer can find ways to reduce the IMD by minimizing the current peaks in the resonators, [17] and [18], and using a higher power handling resonator as the filter's input resonator [19].

## V. Conclusion

A method to reproducibly measure the 3-tone IMD of a superconducting bandpass filter has been demonstrated. The extent to which commercial HTS filters generate 3-tone IMD has been measured. When subject to high out-of-band energy, such as from the transmitter, HTS filters can indeed generate levels of IMD that are high enough to add noise to the base station receiver. For all types of HTS filters studied, the generation of 3-tone IMD varies with the input power with a slope of 1. In most base stations the out-of-band energy is not high enough to cause IMD above the noise floor, but in the example presented, this case is possible. This does not however pose a threat to the commercial deployment of HTS filters as there are design solutions that reduce, or eliminate, the 3-tone IMD.

## ACKNOWLEDGEMENT

The authors had valuable discussions with Scott Bundy, T. Maniwa and Dan Oates. Some of the filters were designed or made by Piotr O. Radzikowski, Jonathan Scupin, Sean Cordone, Nick Lazzaro and David S. Applegate. The reviewers are thanked for their constructive and informed critiques.

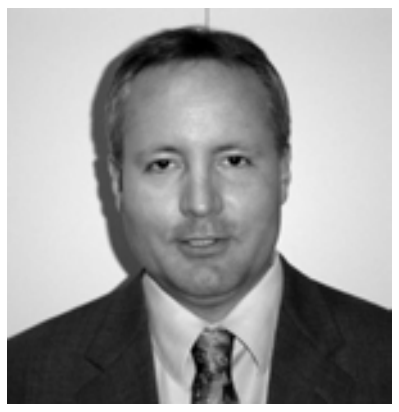

**Stephen K. Remillard** (M'91) was born in Galesburg, IL in 1966. He received the Bachelors degree with honors in physics from Calvin College, Grand Rapids, MI in 1988, an MS in physics and a Ph.D. in physics from the College of William and Mary, Williamsburg, VA in 1990 and 1993.

Part of his doctoral research was done at the Naval Research Laboratory where he studied the microwave properties of high temperature superconductors (HTS). He also researched the fabrication of HTS materials at BUGH Wuppertal in Germany and at Los Alamos National Laboratory. After finishing his Ph.D. he went to Illinois Superconductor Corporation (now ISCO International) where he investigated ways to improve the power handling of HTS RF filters. Since 1998 he was the director of engineering at ISCO International, where he led the development of HTS filter systems. He is now with Agile Devices, Inc in Evanston, IL.

**Huai-ren Yi** (SM'02) received the B.S. degree in physics from University of Science and Technology of China, Hefei, in 1988, the M.S. degree in physics from Chinese Academy of Sciences, Beijing, in 1991, and the Ph.D. degree in physics from Chalmers University of Technology, Gothenburg, Sweden, in 1996.

He was a Staff Member from March 1991 to February 1993 with the Institute of Physics, Chinese Academy of Sciences in Beijing, China; a Research Assistant from March 1993 to November 1996 with the Department of Physics, Chalmers University of Technology, Gothenburg, Sweden; a Post-Doctoral Fellow from December 1996 to August 1998 and a Scientific Employee from September 1998 to October 2001 with the Research Center of Jülich, Germany; a Senior RF Engineer from November 2001 to January 2003 with ISCO International, Inc. Mount Prospect, IL USA; In January 2003, he joined Netcom, Inc. as a Principal Engineer. His fields of interest are the design and realization of high-Tc superconducting thin film filters, dielectric resonator filters and tunable filters. His previous research fields include high-Tc thin films, Josephson junctions and SQUIDs. He has authored/co-authored over 60 papers, five issued patents and two additional patent applications pending.

**Amr Abdelmonem** (M'92 – SM'00) Received the B.Sc. in 1987 and the M.S. in 1990 both in Electrical Engineering from Ain-Shams University, Cairo, Egypt. He earned his Ph.D. in Electrical Engineering from the University of Maryland in 1994.

He is the CEO of ISCO International, which he joined in January 1995 to direct the design and implementation of the first successful field trial of a superconductor filter in a cellular base station. His primary responsibility is defining the optimal direction of the technology and strategic partnership efforts with major 3G OEM wireless equipment providers. Prior to joining ISCO, he held teaching positions at Ain-Shams University and at the University of Maryland. The majority of his research focused on superconducting technology, advanced filter design, semiconductor laser design and optical communication. He has published numerous papers for industry conferences and trade journals. He currently holds five HTS related patents with several additional patent applications pending.